\documentclass[
superscriptaddress,
amsmath,amssymb,
 aps,
twocolumn,
showpacs,
prl,
]{revtex4}
\usepackage{graphicx}
\usepackage{dcolumn}
\usepackage{bm}

\begin{document}

\title{Tidal Waves in ${}^{102}$Pd: A Rotating Condensate of Multiple $d$ bosons}
\author{A.D. Ayangeakaa}
\author{U. Garg}%
\author{M.A. Caprio}
\affiliation{Department of Physics, University of Notre Dame, Notre Dame, IN 46556}%
\author{M.P. Carpenter}
\affiliation{Physics Division, Argonne National Laboratory, Argonne, IL 60439}%
\author{S.S. Ghugre}
\affiliation{UGC-DAE Consortium for Scientific Research, Kolkata Centre, Kolkata 700098, India}
\author{R.V.F. Janssens}
\affiliation{Physics Division, Argonne National Laboratory, Argonne, IL 60439}%
\author{F.G. Kondev}
\affiliation{Nuclear Engineering Division, Argonne National Laboratory, Argonne, IL 60439}%
\author{J.T. Matta}
\affiliation{Department of Physics, University of Notre Dame, Notre Dame, IN 46556}%
\author{S. Mukhopadhyay}
\altaffiliation[Present address: ]{Nuclear Physics Division, Bhabha Atomic Research Centre (BARC), Mumbai 400085, India}
\affiliation{Department of Physics, Mississippi State University, Mississippi State, MS 39762}
\author{D. Patel}
\affiliation{Department of Physics, University of Notre Dame, Notre Dame, IN 46556}%
\author{D. Seweryniak}
\affiliation{Physics Division, Argonne National Laboratory, Argonne, IL 60439}%
\author{J. Sun}
\affiliation{Department of Physics, University of Notre Dame, Notre Dame, IN 46556}
\author{S. Zhu} 
\affiliation{Physics Division, Argonne National Laboratory, Argonne, IL 60439}%
\author{S. Frauendorf}
\affiliation{Department of Physics, University of Notre Dame, Notre Dame, IN 46556}

\date{\today}

\begin{abstract}

Low-lying collective excitations in even-even vibrational and transitional nuclei may be described semi-classically as quadrupole running waves on the surface of the nucleus (``tidal waves''), and the observed vibrational-rotational behavior can be thought of as resulting from a rotating condensate of interacting $d$ bosons. 
These concepts have been investigated by measuring lifetimes of the levels in the yrast band of the $^{102}$Pd nucleus with the Doppler Shift Attenuation Method. 
The extracted $B(E2)$ reduced transition probabilities for the yrast band display a monotonic increase with spin, in agreement with the interpretation based on rotation-induced condensation of aligned $d$ bosons. 
\end{abstract}

\maketitle

\begin{figure}[b]
\hspace*{-0.6cm}
\centering
\includegraphics[scale=1.5]{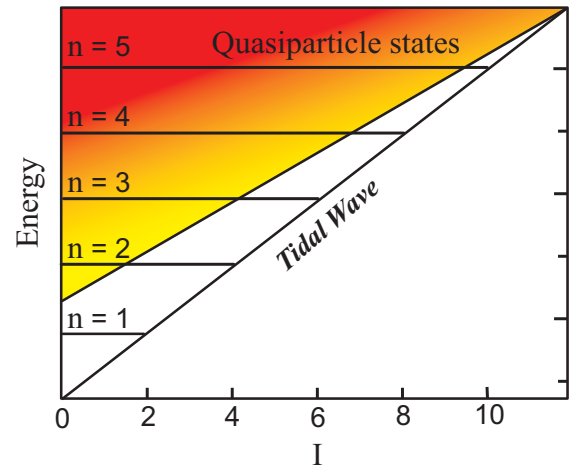}
\vspace*{-0.2cm}
\caption{(Color online) Schematic representation of the location of the collective quadrupole vibrational excitations relative to the quasiparticle excitations. The darker shades approximately indicate higher densities of quasiparticle states. }
\label{fig:vib_qp}
\end{figure}

Collective quadrupole excitations of nuclei are classified as ``rotational'' and ``vibrational'' with the rigid rotor and the harmonic vibrator being the limiting ideal cases. The harmonic vibrations ($d$ bosons) are five-fold degenerate with respect to the angular momentum projection; they  couple to multiplets of different angular momentum. 
 Rotational bands that extend over ten or more states are ubiquitous. Vibrational excitations of spherical nuclei, on the other hand, are much less distinct since the time scale of collective vibrations is not much larger than that of intrinsic quasiparticle excitations. The adiabatic separation of the time scales, which is a prerequisite for the appearance of collective quantum states, rapidly deteriorates with the number of excited quanta (phonons). While evidence for the two-phonon triplet is often observed, identification of all members of the three-phonon multiplet is already problematic (see, for example, Ref.~\cite{Apraham-PRL.59.535}).
 Fig. \ref{fig:vib_qp} schematically shows the location of the vibrational states  and of the quasiparticle excitations for a spherical nucleus. With increasing   phonon number $n$, the collective states are embedded into a progressively dense background of quasiparticle excitations. The coupling  to this quasiparticle background fragments the collective levels and they cease to exist as individual quantum states. The density of quasiparticle excitations is lowest near the yrast line, which is the sequence of states with the lowest excitation energy for a given angular momentum $I$. With increasing phonon number $n$, the yrast members of the vibrational multiplets are expected to keep their identity as collective quantum states the longest. 

In this Letter, we report on the first identification of a seven-phonon yrast state, which can be interpreted as a rotating condensate of seven $d$ bosons that align their spins. The yrast states of $^{102}$Pd are nearly equidistant in energy up to spin $I=16$, which is characteristic of a vibrational sequence. Indeed, the existence of an exceptionally long vibrational band in $^{102}$Pd was already noted by Regan et al.~\cite{Regan-PRL.90.152502}, making this nucleus an attractive case for this investigation. However, in order for these states to represent a sequence of stacked $d$ bosons, the reduced transition probabilities between adjacent states, $B(E2;I\rightarrow I-2)$, must increase linearly with the number of bosons. In this work, lifetime measurements are presented that confirm this expectation. The measurements are complemented by theory, which provides a connection with the underlying microscopic structure that has remained a challenge in the case of multi-phonon excitations.

\begin{figure}
\centering
\includegraphics[scale=0.3]{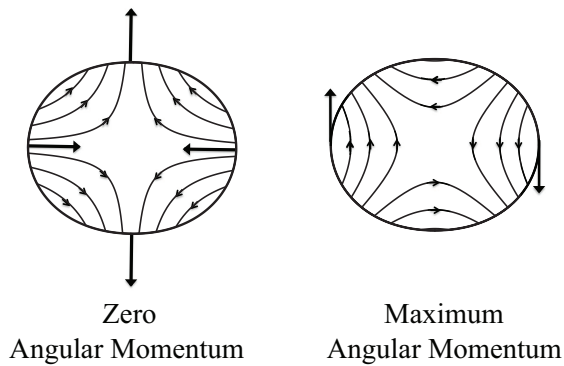}
\caption{Flow pattern of a vibrating droplet of an ideal liquid. The arrows indicate the motion of the surface.
For a given energy, there are degenerate modes, which differ by their  angular momentum. Left: The oscillating motion, which carries no angular momentum. Right: The traveling wave (tidal wave), which carries the maximal angular momentum (yrast state). The outward motion of the surface at south-east and the inward motion at north-east combine such that the eastern crest moves down.}
\label{fig:Flow}
\vspace*{-0.5cm}
\end{figure}

Fig. ~\ref{fig:Flow} illustrates the vibrational modes of minimal and maximal angular momentum  of a classical ideal liquid. The oscillating standing wave on the left-hand side represents the zero-angular momentum members of the vibrational multiplets, which are most rapidly drowned in the sea of quasiparticle excitations. The vibrational yrast states correspond to a wave traveling over the nuclear surface, as illustrated on the right hand side. The surface rotates with the constant angular velocity $\omega$ as is the case for the rotation of a rigid body. However,  the flow pattern is irrotational. As is characteristic of a surface wave,  liquid moves from the wavefront under the crest to the back side. The energy and the angular momentum increase with the amplitude of the wave, but the frequency remains constant. In the case of rigid rotation, the energy and the angular momentum increase with the angular frequency while the shape remains unchanged. The name ``tidal wave'' has been suggested for the yrast mode \cite{Frau10} in analogy to the propagation of ocean tidal waves. The tidal wave then corresponds to a wave packet of aligned $d$ bosons, which rotates with constant angular velocity.   
The tidal wave concept has been
applied previously to near-equidistant band structures of different spin-parity sequences. Specifically, Refs.~\cite{Pattison-PRL.91.182501, Cullen-JPG.31.S1709} interpreted the $\Delta I=1$
sequences of fixed parity in $^{182}$Os as a tidal wave running over a triaxial surface, and  Ref.~\cite{Reviol:2006ly} interpreted the alternating-parity sequences in $^{220}$Th as a reflection-asymmetric tidal wave traveling over a spherical core.

In terms of a rotational interpretation, the difference between a rotor and a tidal wave lies in the way in which angular momentum is generated. A rotor generates angular momentum by increasing the angular frequency $\omega$ at nearly constant deformation (and, hence, constant moment of inertia). A tidal wave nucleus, on the other hand, generates angular momentum by increasing deformation (i.e., changing $\beta, \gamma$) at nearly constant angular frequency $\omega$. For most transitional nuclei, the situation is intermediate: Both mechanisms of generating angular momentum are present, one favored over the other as the angular momentum increases. The rotor-like scenario has been well studied, with the measured reduced transition probabilities, i.e.,  $B(E2)$ values, confirming that the deformation remains rather constant~\cite{Kotlinski1990365}. However, the strong increase of deformation with angular momentum, which characterizes a tidal wave, is demonstrated for the first time in the work reported here.

The mean lifetimes of yrast states in  ${}^{102}$Pd were measured using the Doppler Shift Attenuation Method (DSAM). The experiment was performed at the Argonne Tandem Linear Accelerator System (ATLAS) facility using the ${}^{76}$Ge($^{30}$Si, 4n) reaction, at a beam energy of 110 MeV. The target consisted of 500 $ \mathrm{\mu g /cm^{2}}$ isotopically enriched  ${}^{76}$Ge backed by a 26 mg/cm$^{2}$-thick layer of Au; the backing thickness was sufficient to stop the recoiling nuclei.  A thin layer of 11 $\mathrm{\mu g/cm^{2}}$ Al was sandwiched between the target and backing to prevent the migration of Ge into Au. The emitted $\gamma$ rays were detected with the  Gammasphere array \cite{IYang1990c641} which, at the time of the experiment, comprised ninety eight  Compton-suppressed high-purity Ge detectors, arranged in 16 rings of constant angles relative to the beam direction; a total of 3.1$\times$10${}^{9}$ three- and higher-fold coincidence events were accumulated. The low-spin structure of ${}^{102}$Pd is already well established \cite{Jerrestam1996, Gizon199795, Zamfir-PRC.65.044325}. The present investigation has focused primarily on the measurement of the lifetimes of members of the yrast band. 

The accumulated data were sorted angle-by-angle into several unique-fold event databases, using the BLUE program ~\citep{Cromaz2001519}. For each database, an angle-dependent background subtraction algorithm~\citep{Starosta2003771} was applied, and the resulting, background-subtracted, databases were double gated in order to extract coincidence and error spectra for each ring of Gammasphere for further analysis. 

Doppler-shifted and broadened lineshapes were observed for transitions in the yrast band up the $16^{+}$, 7244-keV level and lifetime analyses were performed using a modified version of the {\small LINESHAPE} analysis package~\cite{Wells1991}. A total of 50000 Monte Carlo simulations of the velocity histories of the recoiling nuclei traversing the target and backing materials were generated in time steps of 0.0089 ps and were subsequently converted into time-dependent velocity profiles. The total number of time steps was set to 198 and the time step value was determined in accordance with the discussion in Ref.~\cite{Wells1991}.  Electronic stopping powers were taken from Ziegler's tabulation~\cite{Ziegler1980}, with low-energy modifications. Energies of in-band transitions were determined from fits to the experimental data and the side-feeding intensities obtained directly from the measured $\gamma$-ray intensities within the yrast band. The side feeding into each level and to the top of the yrast band was modeled by a five-state rotational cascade with independently variable lifetimes. A $\chi^{2}$ minimization for the observed Doppler shifted lineshapes was then performed using the in-band and side-feeding lifetimes, background, and contaminant peak(s) as input parameters. Experimental uncertainties in the extracted lifetimes were determined  based on the behavior of the ${\chi}^{2}$ fit in the vicinity of the minimum~\cite{Chiara2000,lifetimes} by a statistical method using the {\small MINOS}~\cite{James:1975dr} program. Although side-feeding effects were initially included in the analysis, the extracted lifetimes were found to be insensitive to changes in the side-feeding intensities.  More details on the fitting procedure can be found in Refs.~\cite{Chiara2000, Chiara2001}. 

\begin{figure}[b]
\hspace*{-0.4cm}
\includegraphics[scale=0.38]{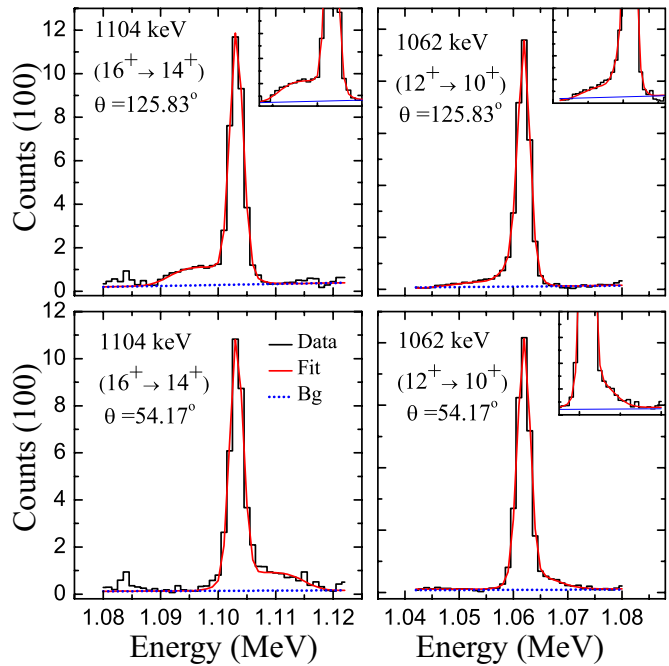}
\caption{\label{fig:dsampic} (Color online) Lineshape fits to the 1104- and 1062-keV transitions in the yrast band in $^{102}$Pd. The upper and lower panels correspond to the backward and forward detectors, respectively. The inserts are expanded views of the lineshapes.}
\end{figure}

Experimental data and lineshape fits at both forward and backward angles for two representative transitions (1104 and 1062 keV) in the yrast band are displayed in Fig.~ \ref{fig:dsampic}. The inserts are expanded views of the corresponding lineshapes. %
The extracted lifetimes of states in the yrast band, the corresponding reduced transition probabilities, and the associated transition quadrupole moments $Q_t$ are summarized in Table \ref{tab:table1}. The reduced transition probabilities were obtained from the measured lifetimes using the expression:
\begin{equation}
B(E2)=\frac{0.0816}{E^{5}_{\gamma}(E2) \tau (E2)} \;   [\mathrm{e^{2}b^{2}}],
\end{equation}
with the $\gamma$-ray energies, $E_{\gamma}$, in MeV, and the partial lifetimes of the transitions, $\tau (E2) $, in ps.

\begin{table}
\caption{\label{tab:table1}%
Measured lifetimes and electromagnetic transition probabilities for the yrast band in ${}^{102}$Pd. Experimental uncertainties were derived from the behavior of $\chi ^{2}$ in the vicinity of the best-fit parameter values.}
\begin{ruledtabular}
\begin{tabular}{cllcc}
\textrm{E${}_{\gamma}$(MeV)}&
\textrm{I${}_{i}^{\pi} \rightarrow$ I${}_{f}^{\pi}$}&
\textrm{$\tau$(ps)}&
\multicolumn{1}{c}{\textrm{$B(E2)$(e${}^{2}$b${}^{2}$)}}&
\multicolumn{1}{c}{\textrm{$Q_{t}$(eb)}}\\
\colrule
0.836 & 6${}^{+}$$\rightarrow$   4${}^{+}$& 1.149(32) & 0.174(22) & 2.36(15)\\
0.902 &  8${}^{+}$$\rightarrow$   6${}^{+}$& 0.720(42) & 0.190(11) & 2.41(07)\\
0.980 &10${}^{+}$$\rightarrow$   8${}^{+}$& 0.409(28) & 0.221(15) & 2.56(09)\\
1.062 &12${}^{+}$$\rightarrow$ 10${}^{+}$& 0.252(18) & 0.240(14) & 2.65(08)\\
1.083 &14${}^{+}$$\rightarrow$ 12${}^{+}$& 0.201(11) & 0.272(15) & 2.80(08)\\
1.104 &16${}^{+}$$\rightarrow$ 14${}^{+}$&0.180(21)\footnote{The lifetime of the 16$^{+}$ level could not be separated from the side-feeding lifetime. The value given is, therefore, the lower limit.} & 0.277(14) & 2.81(07)\\
\end{tabular}
\end{ruledtabular}
\end{table}

\begin{figure}[t]
\includegraphics[scale=0.40]{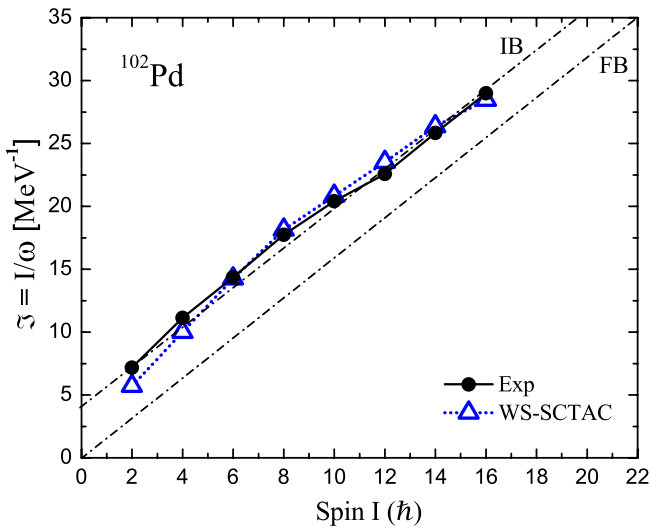}
\caption{\label{fig:j} (Color online) Moments of inertia, ${\cal J}$, as a function of spin, $I$. The dashed line marked FB  is the limit for harmonic bosons. The dashed line marked IB  indicates the near-linear trend of interacting bosons.
WS-SCTAC represents the cranking + micro-macro calculations described in the text.  
}
\end{figure}

\begin{figure}[b]
\includegraphics[scale=0.40]{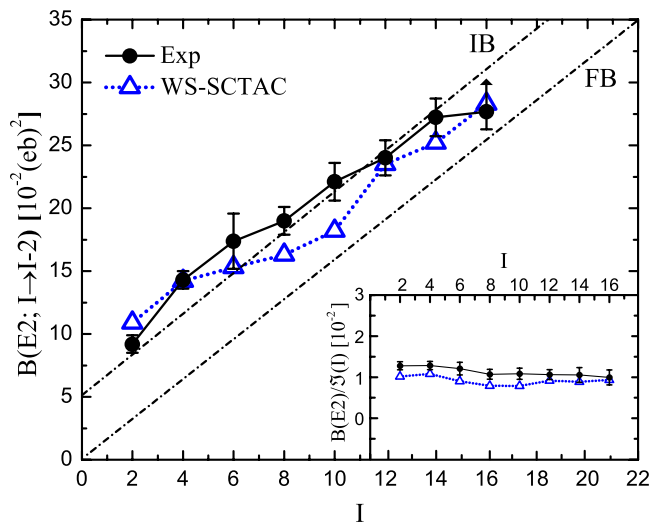}
\caption{\label{fig:be2} (Color online) Reduced electromagnetic transition probabilities for the yrast states of $^{102}$Pd as a function of the spin, $I$. Lifetime data for the 2$^+$ and 4$^+$ states are from Ref.~\cite{DeFrenne20091745}. The dashed line marked FB  is the limit for harmonic bosons. The dashed line marked IB  indicates the near-linear trend of interacting bosons.
WS-SCTAC represents the cranking + micro-macro calculations described in the text. The insert provides the experimental and calculated ratios $B(E2)/\cal J$ as a function of $I$.}
\end{figure}

Fig. \ref{fig:j} provides the experimental moments of inertia ${\cal J}=I/\omega=2I/[E(I)-E(I-2)]$ as a function of spin $I$.  
 The harmonic limit for the free bosons is displayed by the dashed line marked FB and  corresponds to a constant $\omega$ equal to  one half of the vibrational frequency $\Omega$. 
The experimental ${\cal J}$ moment is a nearly linear function of $I$ as indicated by the dashed line marked IB (for interacting bosons).
 It deviates from FB by the small offset at $I=0$, which is a measure of the anharmonicity. The angular momentum increases due to the increase of 
 ${\cal J}$ while $\omega$ remains nearly constant.   
 
 The measured $B(E2)$ values (Fig. \ref{fig:be2}) increase linearly with $I$, 
 such that the ratio $B(E2)/{\cal J}$ is constant within the experimental uncertainties (see the insert). This demonstrates that the angular momentum gain originates from
 the increase of the wave amplitude (deformation) while the rotational frequency does not increase significantly. 
 This is in stark contrast to a typical rotor where the transition probabilities remain relatively constant over a given spin range, while the increase of
 the rotational frequency translates into a gain of angular momentum. This provides the first evidence that the yrast line of $^{102}$Pd
corresponds to a slightly anharmonic tidal wave and the measurements reported here establish the characteristic increase of the tidal wave amplitude up to $I=14$.    
  
It is remarkable that the anharmonicity shows up in a simple way as a constant upshift of the functions ${\cal J}(I)$ and $B(E2; I\rightarrow I-2)$. This 
  may be interpreted as follows: The yrast line of $^{102}$Pd is characterized by an anharmonic wave that starts out with a small deformation (i.e., non-aligned $d$ bosons are present with a small probability)  and increases in amplitude along the sequence of states by adding aligned $d$ bosons. 

The transition quadrupole moment measures the amplitude of the tidal wave. For $^{102}_{46}$Pd$_{56}$, it is 
$Q_{t}=2.8$ eb at $I=14$, to be compared with  $Q_{t}=2.96$ eb for the $I=2$ state in the nucleus $^{110}_{46}$Pd$_{64}$, which is located in the middle of the neutron shell~\cite{NDS-Gurdal20121315}. 
 In the  case of $^{110}$Cd, the experimental amplitude of the tidal wave reaches $Q_t =2.53$ eb  for $I=8$~\cite{NDS-Gurdal20121315}. [All quoted $Q_{t}$ values have been obtained from the lifetime information provided in the cited references.]
  In this case, the $s$ configuration with the rotational aligned pair of $h_{11/2}$ quasineutrons
 becomes yrast for $I\geq12$~\cite{Regan-PRL.90.152502}. Its nearly  constant  smaller deformation ($Q_t \sim 1.7$ eb) thereafter signals the transition to the rotational regime (c.f. \cite{Frau10}). 
 In the case of  $^{102}$Pd, the rotational aligned $s$ configuration does not become yrast until after $I$=16, which, as mentioned previously, is one reason why this nucleus was deemed very attractive from the point of view of studying the boson
 condensate. Unfortunately, it was not possible to determine the lifetimes of these ``rotational'' states from the present data.  
 
The tidal wave concept allows for a semiclassical calculation of the energies and $B(E2)$ values of the yrast states in vibrational and transitional nuclei.
The tidal wave has a static deformed shape in the co-rotating frame of reference.
This has led to the microscopic description suggested in Ref. \cite{Frau10}, which is  based on  the cranking model without resorting to the small amplitude approximation. 
 Ref. \cite{Frau10} applies the shell-correction tilted-axis cranking (SCTAC) version \cite{qptac} of the cranking model to even-even nuclides with $44\le Z\le 48$ and $56\le N\le 66$. 
In this approach, the energy surface is calculated by the micro-macro method, subject to the constraint that the expectation value of the angular momentum operator equals $I$. The energy is minimized with respect to the deformation parameters $\beta$ and $\gamma$.  Deformed solutions are found in Ref. \cite{Frau10} for $I\ge 2$, even when the solution  was spherical for $I=0$.
  These calculations reproduce the collective yrast states rather well. They also 
describe the intrusion of the aligned $h_{11/2}$ two-quasineutron states into the yrast line, which 
causes the backbending phenomenon seen in most of the nuclei studied. 
The details of the present calculations are the same as those of Ref. \cite{Frau10}, with the exception that the modified oscillator potential is replaced by
a deformed Woods-Saxon potential with the so-called ``universal'' parameters \cite{universal}. 
The results are labeled by SCTAC in Figs. \ref{fig:j} and  \ref{fig:be2} (``TAC'' refers only to the code; the 
self-consistent solution actually 
rotates about the intermediate principal axis of the slightly triaxial shape). It should be noted that there are no parameters adjusted to the experiment.

As seen in Fig.~\ref{fig:j}, the observed behavior of ${\cal J}$ as a function of spin $I$ is reproduced well by these calculations. 
The calculated  ${\cal J}(I)$ moment remains close  to the IB line because the ground state
configuration has been employed throughout. 
The calculated  $B(E2;I\rightarrow I-2)$ values (Fig.  \ref{fig:be2}) also follow the experimental data overall, albeit with somewhat larger fluctuations than those exhibited by the experimental values. The reason for this apparent discrepancy lies in the fact that the deformation parameters are those associated with the minimum of the energy calculated by the micro-macro method. This method neglects zero point fluctuations of the shape which, when properly accounted for, are expected to result in washing out of these fluctuations.     
Thus, the semiclassical cranking+micro-macro calculations describe the yrast states of $^{102}$Pd
up to the seven boson state rather well, once the anharmonicities are correctly accounted for. 

In summary, by measuring the lifetimes of the yrast states in  $^{102}$Pd up to a spin of $I=14$, the first clear evidence has been provided for rotational-induced  condensation of aligned $d$ bosons. The linear increase of the reduced transition probability with the boson number $n=I/2$, which is expected for condensation, was observed up to $n$=7. The mutual interaction of bosons causes anharmonicities which appear as a constant upshift of the reduced transition probability and the moment of inertia as function of spin $I$. This upshift may be interpreted as caused by a small fraction of non-aligned $d$ bosons to which the aligned $d$ bosons are added. Semiclassically, the condensate  represents a tidal wave traveling over the nuclear surface with constant  angular velocity (equal to one half of the vibrational frequency), where the angular momentum gain arises from the increase of the wave amplitude. Since the wave motion is not quite harmonic, a slight increase of the rotational frequency with spin is seen.  The tidal wave is described in the framework of the cranking model based on the micro-macro method, which describes the data well $without$ any adjustment of the model parameters.       

We thank R.V. Ribas for providing the DSAM analysis code and C.J. Chiara for helpful discussions about the DSAM analysis procedures. This work has been supported in part by the National Science Foundation (Grant Nos. PHY07-58100 and PHY-1068192), and by the Department of Energy, Office of Nuclear Physics, under Grant Nos. DE-FG02-95ER40934 (UND) and  DE-FG02-95ER40939 (MSU), and Contract No. DE-AC02-06CH11357 (ANL).

\bibliographystyle{apsrev}
\bibliography{tidal}
\end{document}